\documentclass[showpacs,preprintnumbers,amsmath,amssymb]{revtex4}
\makeatletter
\begin{document}
\title{Spin Hall Effect For Anyons}
\author{S. Dhar}
 \email{sarmi_30@rediffmail.com}
 \affiliation{Physics Department, Bankura Sammilani College,\\
  West Bengal, India}
\author{B. Basu}
 \email{banasri@isical.ac.in}
\author{Subir Ghosh}
 \email{subir_ghosh2@rediffmail.com }
 \affiliation{Physics and Applied Mathematics Unit\\
 Indian Statistical Institute\\
 Kolkata-700108, India }

\begin{abstract}
We explain the intrinsic spin Hall effect from generic anyon
dynamics  in the presence of external electromagnetic field. The
free anyon is represented as a spinning particle with an underlying
non-commutative configuration space. The Berry curvature plays a
major role in the analysis.
\end{abstract}
 \pacs{73.43.-f, 72.25.Dc., 03.65.Vf}
\maketitle
\section{Introduction}
In the nascent field of spintronics \cite{wolf} understanding the
dynamics of spin current is extremely important. The prediction
\cite{she} and observation \cite{kato} of an intrinsic Spin Hall
Effect has evoked a lot of interest since the dissipationless Spin
Hall current can be an efficient means of injecting spin current in
(Ga-As) semiconductors. Furthermore the Spin Hall conductivity can
be quantized \cite{bern} and the resulting quantum spin Hall liquid
will have exotic features such as {\it{fractional}} statistics
indicating the presence of {\it{anyons}} \cite{wil} as
quasi-particles. In fact signatures of anyons in Ga-As
heterostructures have been reported \cite{any}. An {\it{anomalous
velocity}} component of electrons, proposed long ago in \cite{lut},
is responsible for Spin Hall \cite{she} and Anomalous Hall Effect
\cite{ahe}. The {\it{semi-classical}} analysis of Bloch electrons in
solid in the presence of external electromagnetic fields, pioneered
by Chang and Niu \cite{ch,bl}, has shown that this anomalous
velocity is induced by  the  intrinsic Berry  curvature \cite{berry}
in Bloch bands \cite{berry1}. This discussion brings out the
perspective of the work reported in this Letter.

The Berry phase emerges during the evolution  of a particle with a
spin by introducing a spin gauge field. Such a particle with a spin
can be described by a vector (multicomponent) wavefunction
\cite{bl}. Anyons possess arbitrary spin \cite{wil} and, conforming
to the above idea, this makes them a prime candidate in Berry phase
study. Indeed, an analogue of the Dirac equation,  to describe a
free relativistic anyon, has been formulated \cite{jn} that requires
an infinite component wavefunction. However, instead of exploiting
the
   multicomponent anyon wavefunction of \cite{jn}, we
will employ   the {\it{spinning particle}} model \cite{cnp,sg,hor1}.
In this model the anyon wavefunction is a {\it{single component
scalar}} and the arbitrary spin is induced by the underlying NC
configuration spacetime of anyon. The NC parameter appears as the
anyon spin.
   The
{\it{non-Abelian}} nature of the $U(1)$ gauge theory in NC spacetime
has already been noticed \cite{nc}. We find its echo in the
condensed matter scenario where the Berry gauge field assumes a
non-Abelian character.

 We study dynamics of anyons in the presence of
external electromagnetic field \cite{cnp,sg,hor1}. Their behavior is
quite different from that
 of a point charge due to the inherent spin-orbit coupling in the former. We interpret
 this system as an effective model of semi-classical dynamics \cite{ch} of anyon excitations
 in the Ga-As alloy.

 We start by representing the general framework of anyon phase space
 and identify the Berry phase. Then we consider a particular anyon
 model mentioned above \cite{cnp,sg,hor1} and obtain the explicit
 forms of the Berry potentials. We show that the Berry phase plays
 an important role to obtain the anomalous velocity and also spin
 Hall conductivity.

 \section{general framework}
We treat (free relativistic) anyons as $2+1$-dimensional spinning
particles and
  consider them in an external electromagnetic field
 ($E_1,E_2,B$). In this configuration the noncommutative geometry
 play the key role of the anyon configuration space (or more generally phase space)
\cite{cnp,sg,hor1} and very interestingly Berry curvature emerges
here through NC geometry. The interesting connection between NC
spacetime  and Berry curvature effect was demonstrated recently in
\cite{ber2}, in the context of momentum space singularity in
anomalous Hall effect \cite{ahe}.

The anyon phase space variables $(r_\mu ,p_\mu )$ are the covariant
physical degrees of freedom. But, as mentioned above, they obey an
NC algebra that reduces to the canonical phase space Poisson
brackets for $s=0$. Because of this $(r_\mu ,p_\mu )$ can not be
used directly in the Einstein-Brillouin-Keller quantization scheme
\cite{ch} or in the identification of the Berry potentials
\cite{bl}. However, one can "solve" the NC algebra in terms of a
{\it{canonical}} (Darboux) set of variables $[R_\mu ,P_\nu
]=-ig_{\mu\nu},~ [R_\mu ,R_\nu ]=[P_\mu ,P_\nu ]=0$. The set $(R_\mu
,P_\mu )$ can be used in the Einstein-Brillouin-Keller framework.
From the mapping between $(r_\mu ,p_\mu )$ and $(R_\mu ,P_\mu )$
the all important  Berry potentials $A^{p}_{\mu}$, $A^{r}_{\mu}$ can
be simply read off \cite{bl}. However, to compare with experiments,
we have to finally re-express the results in terms of the physical
degrees of freedom $(r_\mu ,p_\mu )$. Explicitly, the Darboux
transformation, Berry potential and curvature components
$\Omega^{rp}_{\mu\nu}$ etc. are defined as in \cite{bl},
\begin{equation}
R_{\mu}\equiv r_\mu -A^{p}_{\mu}~,~P_{\mu}\equiv p_\mu
+A^{r}_{\mu};~~\Omega^{rp}_{\mu\nu}=[p_\mu, A_\nu ^p]-[A_\mu
^r,r_\nu ]+[A_\mu ^r,A_\nu ^p]. \label{be}
\end{equation}
The other curvature components follow in an obvious way.  Notice
that the NC phase space requires a more general definition of the
curvature $\Omega^{rp}_{\mu\nu}$ than the one prescribed in
\cite{bl}, apart from the non-abelian  $[A^r,A^p]$ commutator term.
The Hamiltonian equations of motion are,
\begin{equation}
\dot p_\mu =-i[p_\mu ,G]~;~~\dot r_\mu =-i[r_\mu ,G], \label{eq}
\end{equation}
where $G$ is generator of (relativistic) time evolution.

The Einstein-Brillouin-Keller quantization condition is
straightforward in terms of canonical $(R_\mu ,P_\mu )$ coordinates:
\begin{equation}
\oint \vec P. d\vec R =j+\frac{\nu}{4} \label{ebk},
\end{equation}
where $j$ is an integer and $\nu$ the Maslov index.
 In terms of $(r_\mu, p_\mu )$, the physical variables, (\ref{ebk}) read,
\begin{equation}
 \oint \vec P .d\vec R =\oint (\vec p +\vec A^{r}).d(\vec r
-\vec A^{p})= \oint \vec p. d\vec r +\oint (\vec A^{r}.d\vec r +\vec
A^{p}.d\vec p ), \label{eq1}
\end{equation}
where geometric part of $\varphi =\oint (\vec A^{r}.d\vec r +\vec
A^{p}.d\vec p )$ is generally termed as the Berry phase.

\section{Anyon model}

 After discussing the general setup we now turn to the particular
case at hand where the anyon wavefunction is a single component
scalar and the arbitrary spin is induced by the NC parameter. We
will restrict ourselves to the lowest non-trivial order in the
electromagnetic coupling $e$ and consider a generalized anyon model
\cite{hor1} with arbitrary gyromagnetic ratio $g$. In these anyon
models \cite{cnp,sg,hor1} the spin tensor $S_{\mu\nu}$ is not
independent, $S_{\mu\nu}=s\epsilon_{\mu\nu\lambda}p_\lambda
/\sqrt{p^2}$ where $s$ and $p_\mu $ are the arbitrary spin parameter
and momentum respectively. Hence spin operators can always be
replaced by momentum operators and spin effects, {\it{e.g.}} in
equations of motion, derived using the NC phase space algebra), are
identified through the parameter $s$.

 The
anyon dynamics is governed by the following generator $G$
 \cite{hor1} and $O(e)$ NC phase space brackets
\cite{cnp,sg,hor1}:
\begin{equation}
G=-\frac{1}{2m}(p^2-m^2+\frac{ge}{2}S_{\mu\nu}
F^{\mu\nu})=-\frac{1}{2m}[p^2-\{m^2-\frac{ges}{p}(\vec p\times \vec
E + Bp_0)\}],\label{4}
\end{equation}
\begin{equation}
[r_{\mu},r_{\nu}]= if_{\mu\nu}-ie(fFf)_{\mu\nu}~;~
[p_{\mu},r_{\nu}]=ig_{\mu\nu}-i(Ff)_{\mu\nu}~; ~
[p_{\mu},p_{\nu}]=ieF_{\mu\nu}, \label{3}
\end{equation}
where
$f_{\mu\nu}=s\epsilon_{\mu\nu\sigma}p^{\sigma}/(p^2)^{\frac{3}{2}}$,
the metric is $g_{00}=-g_{ii}=1$ and
$F_{i0}=E_i,F_{ij}=\epsilon_{ij}B$. With this algebra (\ref{3}), the
Lorentz generator $J_\mu $ that transforms $r_\mu ,p_\mu $
correctly, contains a spin-part \cite{cnp,sg},
\begin{equation}
J_\mu =\epsilon_{\mu\nu\lambda}r^{\nu}p^{\lambda}+sp_\mu /p ,
\label{j}
\end{equation}
and is structurally very similar to the angular momentum defined by
Murakami et.al. in  \cite{ahe}.

The canonical coordinates $(R_\mu ,P_\mu )$ to $O(e)$ are computed
in terms of the physical $(r_\mu ,p_\mu )$ variables, by exploiting
the relations derived in \cite{cnp} and we obtain,
$$
P_{\mu}=p_{\mu}+\frac{e}{2}F_{\mu\nu}(r^\nu +s\alpha^\nu[p])\equiv
p_\mu +A^{r}_{\mu},$$
\begin{equation}
R_{\mu}=r_{\mu}-s\alpha_{\mu}[\tilde
p]-\frac{es}{2}F_{\rho\nu}\{(r^{\rho}-s\alpha^{\rho}
[p])\frac{\partial\alpha_{\mu}}{\partial
p_{\nu}}+s\alpha^{\rho}\frac{\partial\alpha^{\nu}}{\partial
p^{\mu}}\}\equiv r_\mu -A^{p}_{\mu} \label{2}
\end{equation}
where
$$\alpha_{\mu}[p]=(\epsilon_{\mu\nu\rho}p^{\nu}\eta^{\rho})/\lambda
,~~ \lambda =p^2+\sqrt{p^2}p_0,~~\eta_{\mu}=\{1,0,0\},$$
$$\tilde p_\mu =p_\mu
+\frac{e}{2}F_{\mu\nu}(r^{\nu}+s\alpha^{\nu}[p])$$ and the Berry
potentials are introduced. Explicit forms of the potentials, to
$O(e)$, are,
\begin{equation}
A^{r}_{0}=\frac{e}{2}(\vec E.\vec r
-\frac{s}{\lambda}\epsilon_{ij}E_ip_j )
~,~A^{r}_{i}=\frac{e}{2}(r_0E_i-B(\epsilon_{ij}r_{j}+\frac{s}{\lambda}p_i))
, \label{5}
\end{equation}
\begin{equation}
A^{p}_{0}=0~,~A^{p}_{i}=-\frac{s}{\lambda}(1+\frac{esB}{2\lambda})\epsilon_{ij}p_j
. \label{6}
\end{equation}
Thus the Berry curvatures induce the NC brackets (\ref{3}):
$$
[r_i,r_j]=i\frac{sp_0}{p^3}[\epsilon_{ij}+\frac{es}{p^3}(p_iE_j-p_jE_i+p_0B\epsilon_{ij})]\equiv
i\Omega^{pp}_{ij}~;~[p_i,p_j]=ieB\epsilon_{ij}\equiv
-i\Omega^{rr}_{ij},$$
\begin{equation} [p_i,r_j]=-i\delta_{ij}+i\frac{es}{p^3}(\epsilon_{jk}E_ip_k-Bp_0\delta_{ij})\equiv i\delta_{ij}- i\Omega^{rp}_{ij},~~etc.
\label{7}
\end{equation}
It should be noted that external $F_{\mu\nu }$ are also included in
our definition of the curvature $\Omega_{\mu\nu }$.   $e=0$ and
$es$-terms yield  the purely intrinsic part and spin-dependent part
respectively in the curvature. Clearly, even to $O(e)$, the
commutator terms $[A_\mu,A_\nu]$ are non-zero which shows the
non-Abelian nature of the curvature like $U(1)$ gauge theory in NC
space time \cite{nc}. In our formalism, this is due to the
underlying NC geometry (\ref{3},\ref{7}).

Next we derive the equations of motion (see also \cite{hor1}) by
using (\ref{eq},\ref{4},\ref{7}):
\begin{equation}
\dot r_i=\frac{p_i}{m}+(1-\frac{g}{2})\frac{es}{mp}[\epsilon_{ij}E_j
+(p_0B+\vec p\times\vec E)\frac{p_i}{p^2})],~ ~\dot
p_i=\frac{e}{m}(-p_0E_i+B\epsilon_{ij}p_j). \label{8}
\end{equation}
Notice that in the {\it{anomalous}} (or non-canonical) part of the
velocity equation for $\dot r_i $, the $g$-term comes from the
Hamiltonian and the rest is contributed by the coordinate-momentum
mixed curvature $\Omega^{rp}$ whereas the canonical Lorentz force
equation for $\dot p_i $ is induced by $\Omega^{rr}$.\\

\section{physical significance }

 In the Berry phase contribution, the intrinsic
term (in the non-relativistic limit),
\begin{equation}
\varphi \mid _{e=0}=\oint \vec A^p(e=0) d\vec p =\frac{s}{m^2}\times
area~ of~ p-orbit
 \label{bb1}
\end{equation}
will modify  the energy spectra \cite{ch} and density of states
\cite{berry1} of the excitation. Indeed, there are further
spin-orbit ($es$) contributions in $\varphi$ as well, that can be
generated from (\ref{5},\ref{6}). The equations of motion (\ref{8})
explicitly shows the effect of spin-orbit coupling. The "exotic"
particle model studied in \cite{hor1,horv}, which is  a
generalization of the anyon with an anomalous gyromagnetic ratio,
also has a similar spin orbit interaction \cite{horv}.

From the equations of motion, in the non-relativistic limit, we get,
\begin{equation}
\dot r_i=[1+(1-\frac{g}{2})\frac{es}{m^3}\vec p\times \vec
E]\frac{p_i}{m}+(1-\frac{g}{2})\frac{s}{m^2}\epsilon_{ij}\dot p_j
 \label{a1}
\end{equation}
where the last term is the anomalous velocity component. So anyon
dynamics in the presence of external electromagnetic field naturally
yields the anomalous velocity and the Berry curvature in
{\it{mixed}} position-momentum space, (first highlighted in
\cite{zn}), plays an important role in determining the
{\it{anomalous}} velocity.

On the other hand, coming to the Hall effect considerations, we
rewrite the velocity equation in (\ref{a1}) in the form,
\begin{equation}
\dot
r_i=[1+(1-\frac{g}{2})\frac{esB}{m^2}]\frac{p_i}{m}+(1-\frac{g}{2})\frac{es}{m^2}\epsilon_{ij}E_j
\equiv \frac{p_i}{m^*}+(1-\frac{g}{2})\frac{es}{m^2}\epsilon_{ij}E_j
\equiv,
 \label{a2}
 \end{equation}
 where the $O(1/m^4)$ term is dropped.
The effective  mass is now $m^*$ and the last term signifies Hall
motion since it induces
 a velocity, transverse to the Electric field. For $\vec E=(E_x,0)$ we find
\begin{equation}
\dot x=\frac{p_x}{m^*},~~\dot
y=\frac{p_y}{m^*}-(1-\frac{g}{2})\frac{es}{m^2}E_x.
 \label{a3}
\end{equation}
Clearly the Hall conductivity is given by \cite{ezawa}
\begin{equation}
\sigma_{xy}=j_y/E_x =(e\dot
y)/E_x=(1-\frac{g}{2})\frac{se^2}{m^2}\equiv (e^*)^2 \nu .
\label{a4}
\end{equation}
Here, $e^*$ can be considered as the effective (or fractional)
charge. The parameter $\nu $ is the intrinsic Berry phase
(\ref{bb1}) and as expected it also appears in the intrinsic NC
coordinate algebra: $[x_1,x_2]=is/m^2\sim i\nu$ and following
\cite{ezawa} it  can be identified as the "magnetic length". Hence,
as we set out to demonstrate,{\it{ the Hall conductivity
$\sigma_{xy}$ is given by the Berry curvature}}.

 Similarly we can estimate the spin Hall conductivity to be,
\begin{equation}
\sigma^{s}_{xy}=J_y/E_x =(s\frac{\left|\vec p\right|}{m}\dot
y)/E_x=\sqrt{1-\frac{g}{2}}e^*s\frac{\left|\vec p\right|}{m} \nu .
\label{a5}
\end{equation}
Typically, $\left|\vec p\right|$ appears as the Fermi momentum (see
Murakami et.al. in \cite{she}).

From the experimental observations of fractional charge and Hall
conductivity, it is possible to obtain values of $s$ and $g$ which
are anyon parameters. This is a new observation.

\section{discussion}

Presence of fractional statistics has been predicted for quantum
Spin Hall liquid and anyon excitations have been observed
experimentally in some semiconductors. It seems natural to study the
effects from anyon dynamics point of view. In
the present work, following Horvathy et. al[18], we have  shown that the anomalous velocity of Bloch
electron in a semiclassical analysis, (resulting in Spin Hall
effect), emerges naturally when equations of motion for anyons are
studied in external electromagnetic field. A Non-Commutative phase
space structure governs the anyon dynamics. In both the frameworks
Berry curvature plays a key role. We have computed the explicit forms of the Berry potentials adopting the method suggested in [10] and also in [18]. The analysis indicates that there is a possible connection  between our results and physical properties of Ga-As alloy
such as anomalous Hall conductivity. There is a natural and
consistent relation between the parameters of our model ({\it{e.g.}}
spin, mass and gyromagnetic ratio) with those of the bulk system
({\it{e.g.}} magnetic length, filling fraction) \cite{ezawa}.
Our interpretation of anyon parameters can suggest connection with condensed matter  systems qualitatively but more more work is needed for quantitative estimates.
Possibility of measuring the anyon spin and gyromagnetic ratio from experimental observations is a new prediction of our scheme.

{\bf{Acknowledgements:}} We thank Peter Horvathy and Diptiman Sen
for correspondence. 
\end{document}